\documentclass[a4paper,11.5pt]{article}
\pdfoutput=1
\usepackage{graphicx}
\usepackage{epstopdf}
\usepackage{amssymb,amsmath}
\usepackage[colon,authoryear]{natbib}
\usepackage{multirow}
\usepackage{booktabs}
\usepackage{adjustbox}
\title{Long-term orbital stability of exosolar planetary systems with highly eccentric orbits}

\author{Kyriaki I. Antoniadou$^1$ and George Voyatzis$^2$\\
\small{Section of Astrophysics, Astronomy and Mechanics}, \\ \small{Department of Physics, Aristotle University of Thessaloniki,} \\
\small{Thessaloniki, 54124, Greece}\\
\small{email: $^1$kyant@auth.gr, $^2$voyatzis@auth.gr}}

\date{\vspace{-5ex}}

\begin{document}

\maketitle
\begin{abstract}
Nowadays, many extrasolar planetary systems possessing at least one planet on a highly eccentric orbit have been discovered. In this work, we study the possible long-term stability of such systems. We consider the general three body problem as our model.  Highly eccentric orbits are out of  the Hill stability regions. However, mean motion resonances can provide phase protection and orbits with long-term stability exist. We construct maps of dynamical stability based on the computation of chaotic indicators and we figure out regions in phase space, where the long-term stability is guaranteed. We focus on regions where at least one planet is highly eccentric and attempt to associate them with the existence of stable periodic orbits. The values of the orbital elements, which are derived from observational data, are often given with very large deviations. Generally, phase space regions of high eccentricities are narrow and thus, our dynamical analysis may restrict considerably the valid domain of the system's location.
\end{abstract} 

{\bf keywords} dynamical stability, periodic orbits, planetary systems.\vspace{.5em}

\begin{figure}[H]
\begin{center}\begin{adjustbox}{width=1\textwidth}
$\begin{array}{@{\hspace{-.8em}}cc@{\hspace{-.3em}}c}
\includegraphics[width=4.55cm,height=4.65cm]{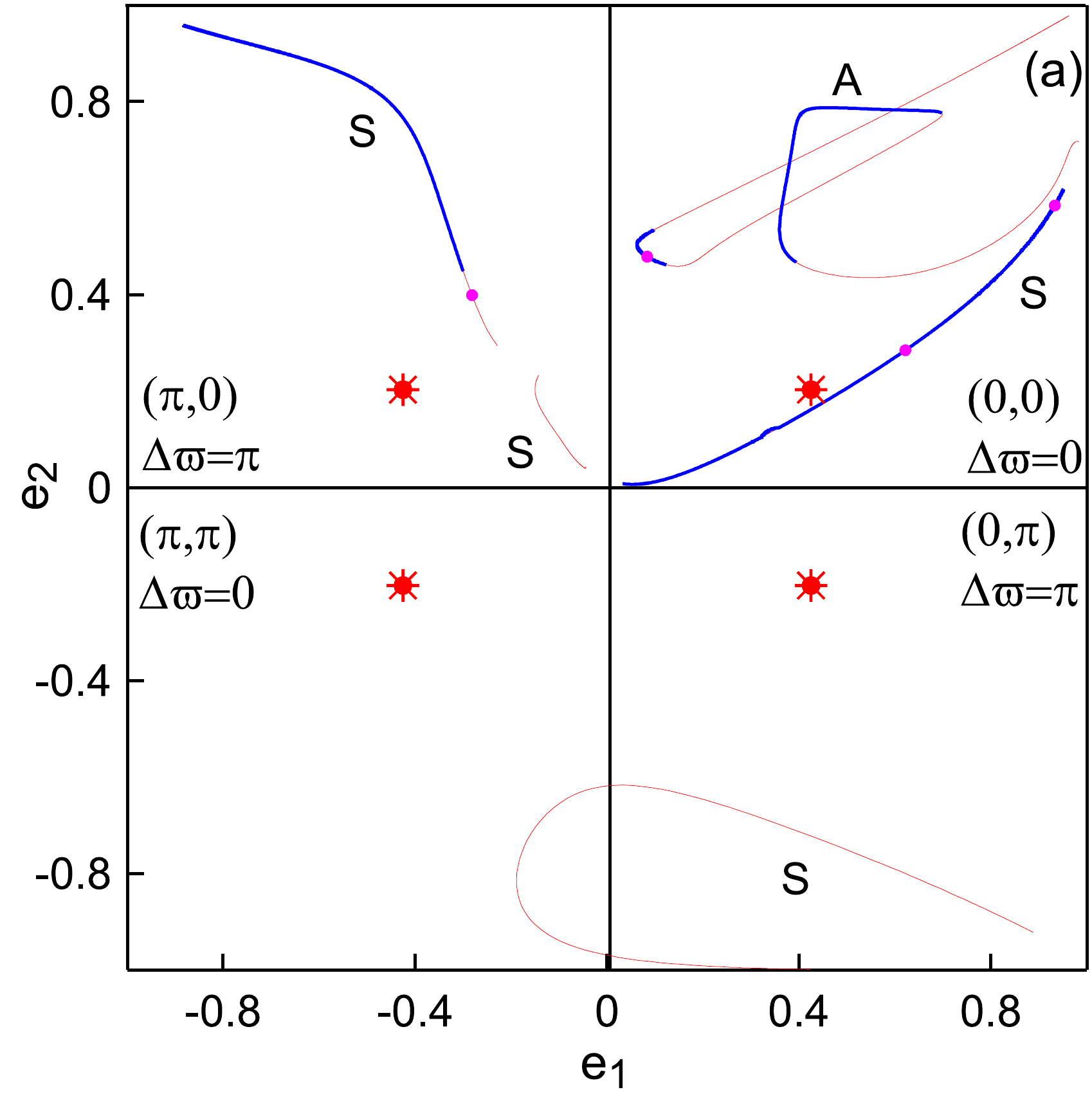}&
\includegraphics[width=4.5cm,height=4.55cm]{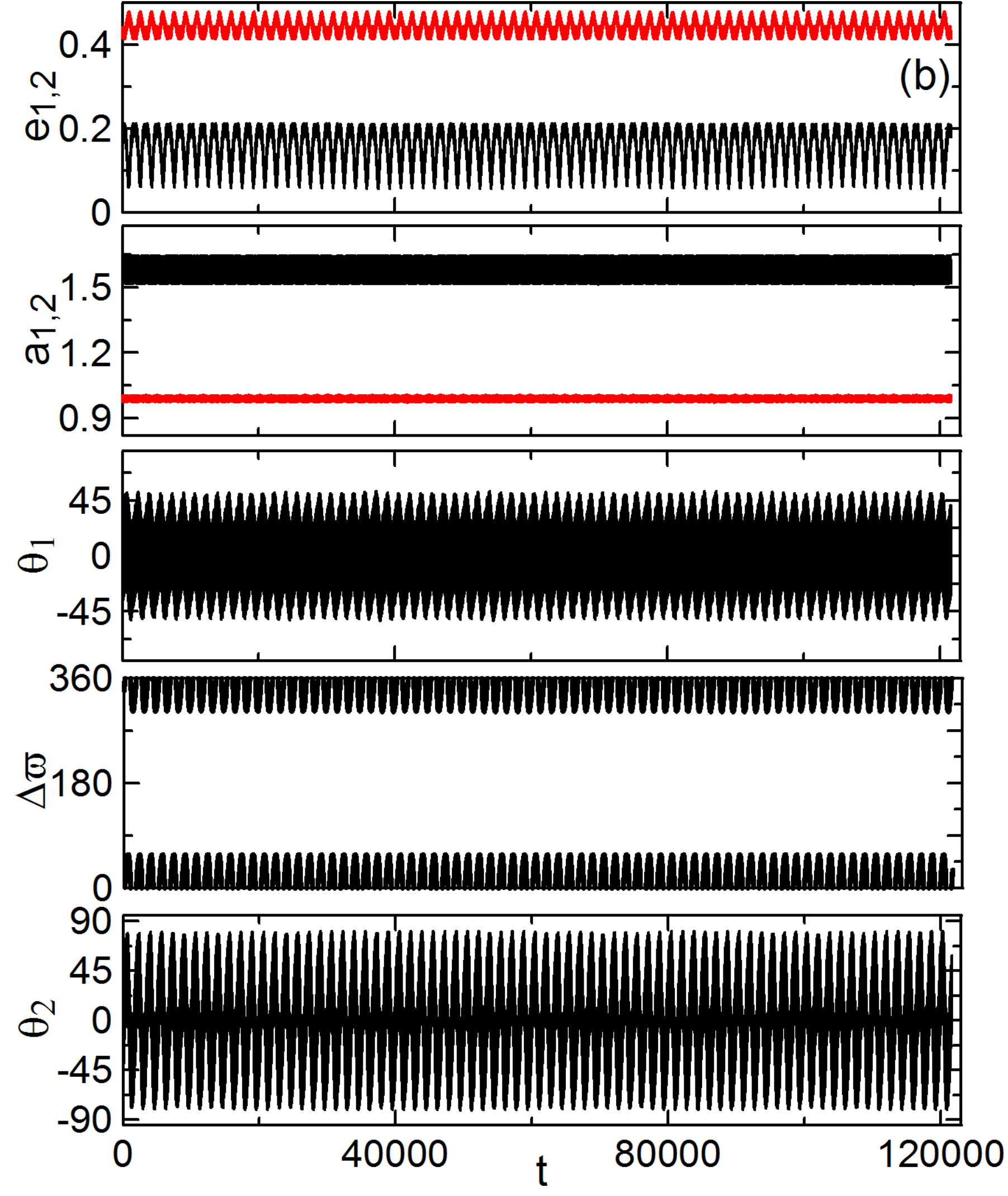}  &
\includegraphics[width=4.25cm,height=4.55cm]{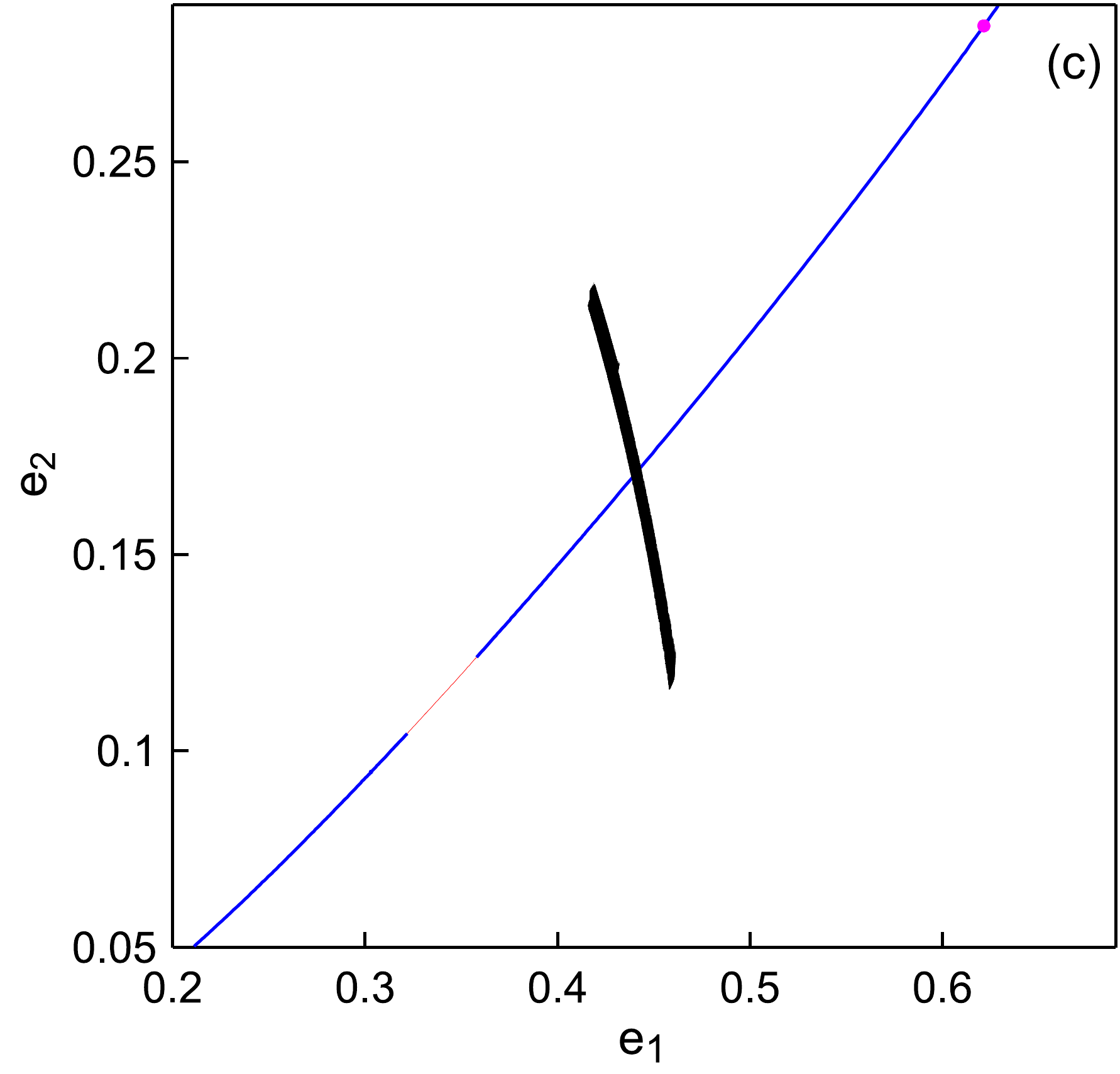} \vspace{-.4em} \\
\vspace{-1.5em} 
\end{array} $\end{adjustbox}
\end{center}
\caption{\textbf{(a)} Families of symmetric (S) and asymmetric (A) periodic orbits in 2/1 MMR for the planetary mass ratio of HD 82943. Blue (red) segments indicate stable (unstable) orbits. Red stars depict $e_{b,c}$, while magenta dots the \textit{vco}. \textbf{(b)} Evolution of orbital elements and resonant angles. \textbf{(c)} Projection of the trajectory on the eccentricities plane and the family of periodic orbits.}\vspace{-0.2em} 
\label{fig1}
\end{figure}

With a vast amount of multiple extrasolar systems  being discovered over the last two decades, we intend to address the following question: How can highly eccentric resonant exoplanets survive? A dynamical mechanism is offered by mean motion resonances (MMR), which can provide phase protection. Our study is based on this mechanism and aims to show its  efficiency. Our methodology breaks down in two steps:
\begin{enumerate}
	\item We compute families of periodic orbits in the planar general three body problem,  which indicate the exact position of a MMR (e.g. \cite{Hadj06}, \cite{av13}). We classify them in different configurations (e.g. \cite{av14}) and compute their linear horizontal and vertical stability. The vertical critical orbits (\textit{vco}) represent the bifurcation points that generate spatial periodic orbits (see Fig. \ref{fig1}a). We justify the MMR and symmetric or asymmetric configuration via librations of resonant angles and apsidal difference about particular angles (Fig. \ref{fig1}b) and depict the planetary evolution (Fig. \ref{fig1}c).
	\item We construct maps of dynamical stability by creating grids defined by two variables (keep the rest orbital elements fixed) and compute the \textit{de-trended} FLI (\cite{gv}) for $t_{max}=250$Ky and finally, we visualize its value. The de-trended FLI takes either small values (visualized by dark colors) suggesting regular motion, or very large values (visualized by yellow color) indicating chaotic motion. The distinction between these cases is very sharp when the integration time of orbits is sufficiently large. Certainly, regular orbits guarantee long-term stability of planetary orbits.      
\end{enumerate}

In the following figures, we present results for the 2/1 resonant dynamics of the two-planet system HD 82943b,c (see also  \cite{tan,bb14}). Maps are computed for the symmetric configuration, which is defined by the resonant angles $\theta_i=2\lambda_2-\lambda_1-\varpi_i=0$, $i=1,2$, where $\lambda_i$ are the mean longitudes and $\varpi_i$ the longitudes of pericenter.  Our analysis shows that the system is located in a region of dynamical stability, which is expanded around a stable periodic orbit. This periodic orbit is vertically stable, thus we may conclude that introducing small mutual inclination the planetary orbits will remain stable. 

We have applied the above methodology to the extrasolar systems HD 3651, HD 7449, HD 89744 and HD 102272 and a complete paper is in preparation.  

\begin{table}
\caption{Data for the  planets HD 82943b,c from \cite{tan}.}\begin{adjustbox}{width=1\textwidth}
\begin{tabular}[htb]{c|cccccc}
\toprule
Planets & \hspace{1em} Mass($M_{Jup}$) & \hspace{1em} a (AU) &  \hspace{1em} e & \hspace{1em} i (deg) & \hspace{1em} $\varpi$ (deg) & \hspace{1em} M (deg)\\
\midrule
HD 82943 c &\hspace{1em} 4.78 & \hspace{1em}1.000 & \hspace{1em}0.425 & \hspace{1em} 19.4 & \hspace{1em} 133& \hspace{1em} 256 \\ 
HD 82943 b &\hspace{1em} 4.8 & \hspace{1em}1.5951 & \hspace{1em}0.203 & \hspace{1em} 19.4 & \hspace{1em} 107& \hspace{1em} 333 \\
\bottomrule\end{tabular}\vspace{-.4em} 
\label{tab1}\end{adjustbox}
\end{table}

\begin{figure}[H]\vspace{-.2cm}
\begin{center}\begin{adjustbox}{width=1\textwidth}
$\begin{array}{@{\hspace{-0.7em}}c@{\hspace{-0.005em}}c@{\hspace{-0.001em}}c}
\includegraphics[width=4.45cm,height=4.45cm]{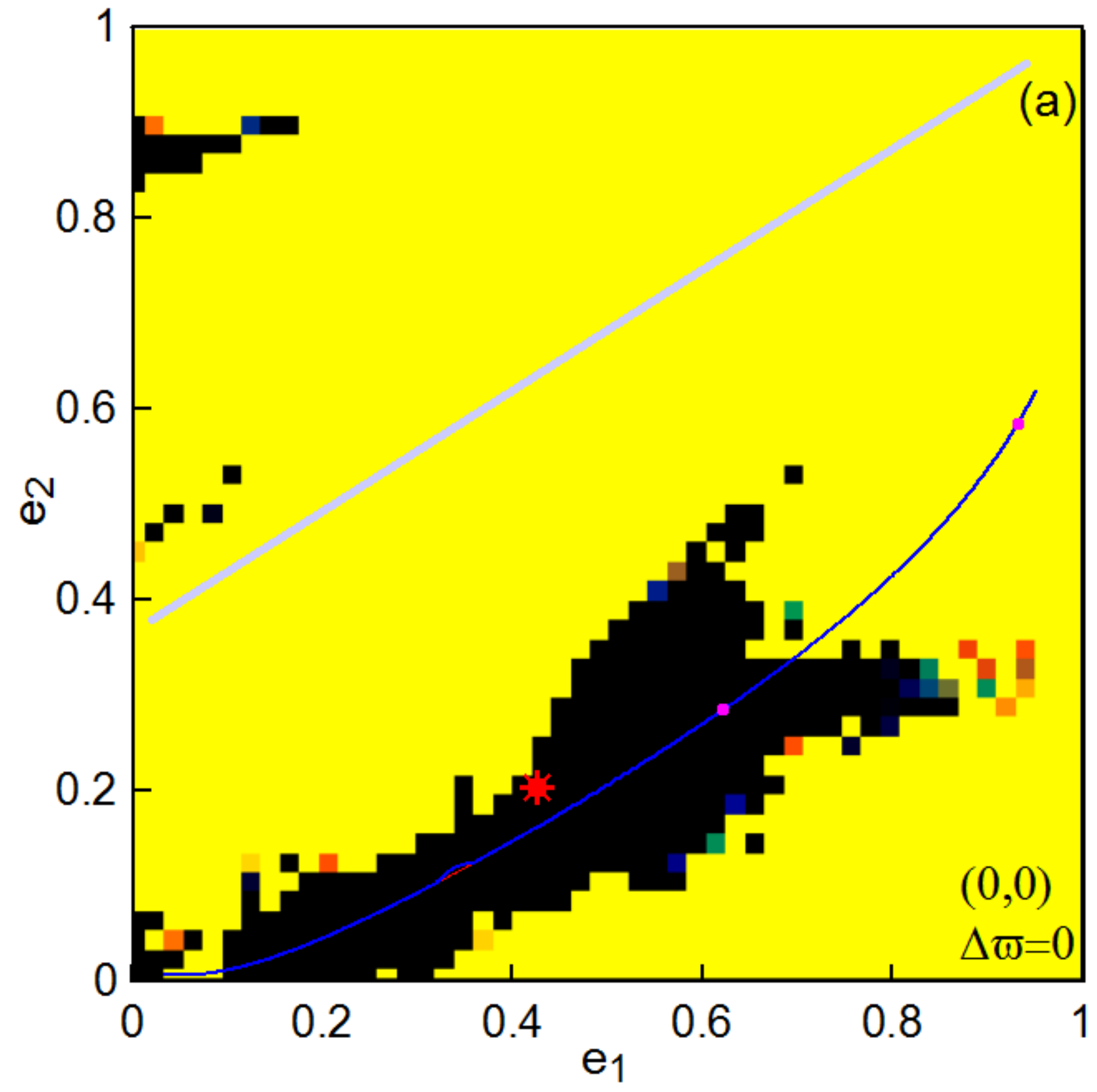}  &
\includegraphics[width=4.45cm,height=4.45cm]{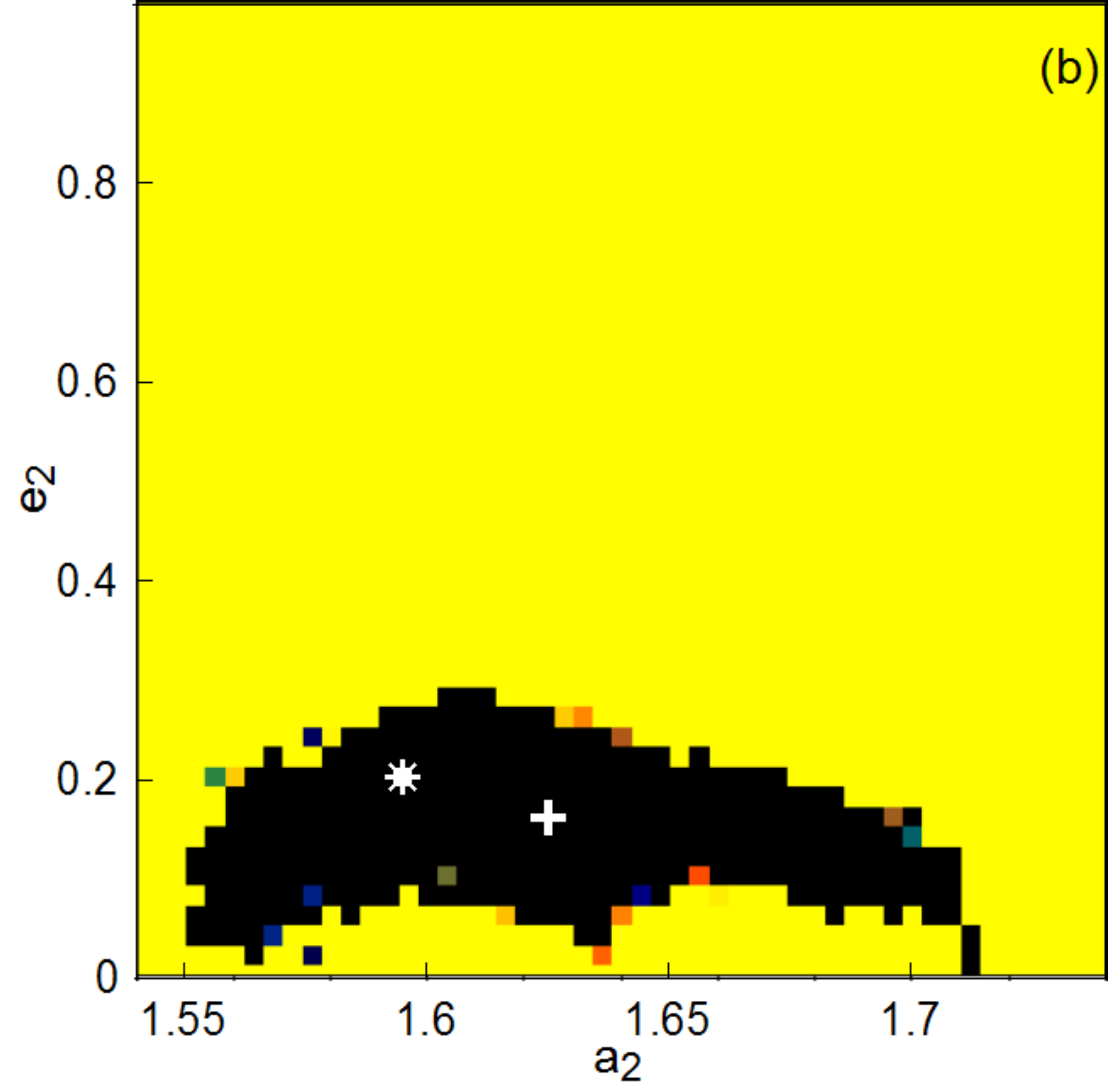}  &
\includegraphics[width=4.45cm,height=4.45cm]{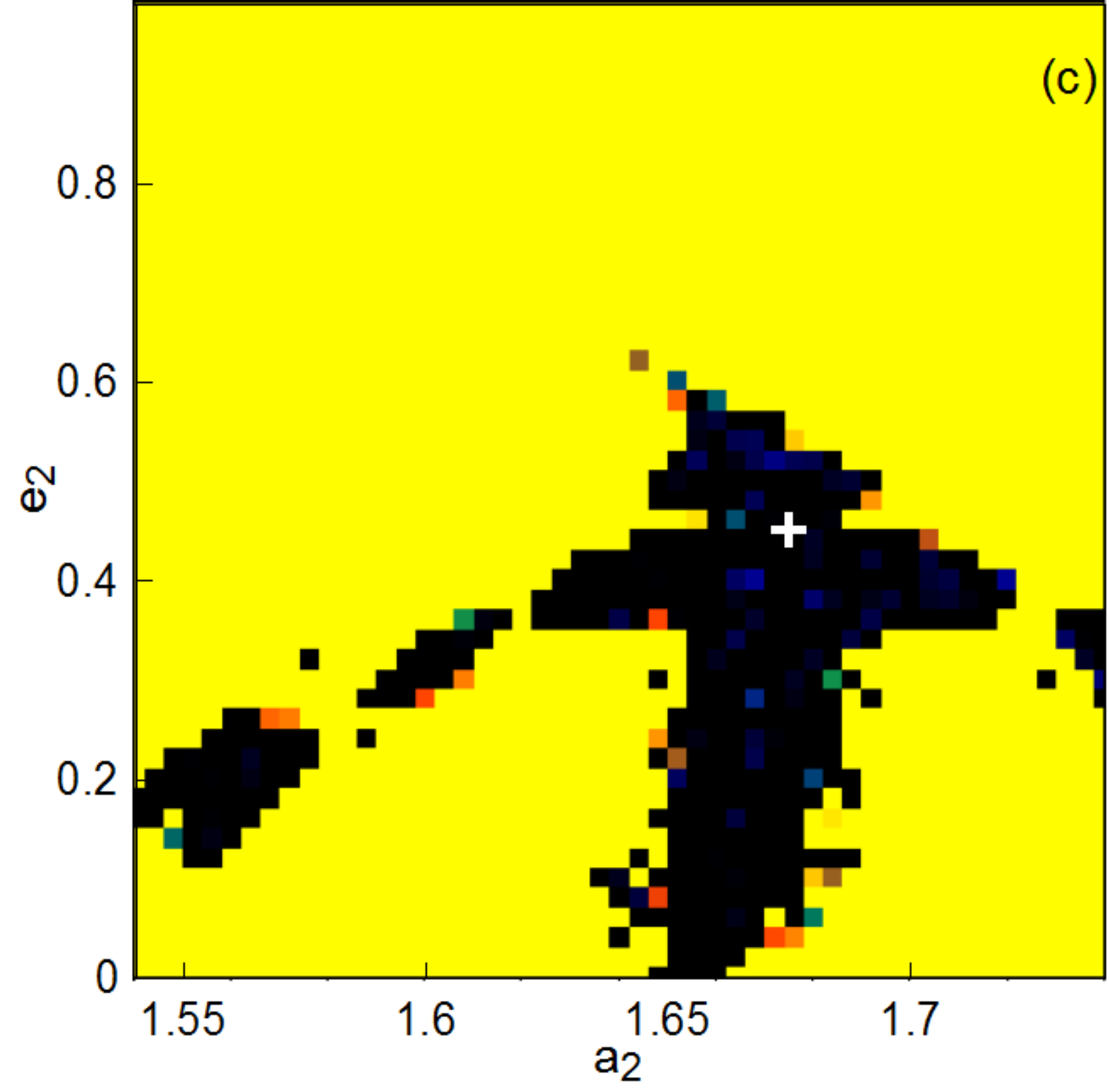} \vspace{-.4em}  \\
\vspace{-1.5em} 
\end{array} $\end{adjustbox}
\end{center}
\caption{ Maps of dynamical stability on: \textbf{(a)} grid plane ($e_1, e_2$) for $a_2$=1.595, where the family of periodic orbits and the collision line (bold gray curve) are also presented, \textbf{(b)} $(a_2, e_2)$ grid plane for $e_1=0.425$, where the resonant periodic orbit is indicated by the cross symbol and the  location of HD 82943b,c by a star and \textbf{(c)} as previously, but for $e_1=0.8$. In all cases we use the normalized value $a_1=1$.} \vspace{-.75em} 
\label{fig2}
\end{figure}

\end{document}